\documentstyle[aps,prb,epsfig]{revtex}
\begin{document}
\twocolumn \wideabs{
\title{Dimensional Crossover in 2D Crossbars}
\author{ K. Kikoin, I. Kuzmenko, S. Gredeskul,  Y. Avishai}
\address{Department of Physics, Ben-Gurion University of the Negev,
Beer-Sheva}
\date{\today}
\maketitle
\begin{abstract}
Spectrum of boson fields and two-point correlators are analyzed in
quantum crossbars (QCB, a superlattice formed by two crossed interacting
arrays of quantum wires), with short range inter-wire interaction.
It is shown that the standard bosonization procedure is valid,
and the system behaves as a sliding Luttinger
liquid in the infrared limit, but the high frequency spectral and
correlation characteristics have either  1D or 2D nature depending
on the direction of the wave vector in the 2D Brillouin zone.
As a result, the crossover from 1D to 2D regime may be
experimentally observed. Plasmon
propagation in arbitrary direction is possible. Periodic energy transfer
between arrays ("Rabi oscillations") is predicted.
\end{abstract}}
\vspace{\baselineskip}
\section{Introduction. From quantum wires to quantum crossbars}

The behavior of electrons in arrays of $1D$ quantum wires was
recognized as a challenging problem soon after the consistent theory
of elementary excitations and correlations in a Luttinger liquid (LL)
of interacting electrons in one dimension was formulated (see
\cite{Voit} for a review). One of the fascinating challenges existing
in this field is a search for LL features in higher dimensions
\cite{Anders}. Although the Fermi
liquid state seems to be rather robust for $D>1$, the possible
way to retain some $1D$ excitation modes in $2D$ and even $3D$
systems is to consider highly anisotropic objects, in which the electron
motion is spatially confined in major part of the real space
(e.g., it is confined to separate linear regions by potential relief). One
may hope that in this case weak enough perturbation does not violate
the generic long-wave properties of the LL state.
Arrays of interacting quantum wires may be formed in organic materials
and in striped phases of doped transition metal oxides.
Artificially fabricated structures with controllable configurations of
arrays and variable interactions are available now due to
recent  achievements in nanotechnology (see, e.g., \cite{Rueckes}).

We start with a discussion of an array of parallel quantum wires.  The
conventional LL regime in a $1D$ quantum wire is characterised by
bosonic fields describing charge and spin modes.  We confine our
discussion to the charge sector (LL in the spin-gapped phase).  The
Hamiltonian of an isolated quantum wire may then be represented in a
canonical form
\begin{equation}
H = \frac{\hbar v}{2}\int\limits_{-L/2}^{L/2} {dx}
\left\{g{\pi}^{2}(x)+
\frac{1}{g}({\partial}_{x}\theta(x))^2\right\}.
\label{D1}
\end{equation}
Here $L$ is the wire length, $v$ is the Fermi velocity, $\theta,\pi$
are the conventional canonically conjugate boson fields and $g$ is the
dimensionless parameter which describes the strength of the interaction
within the chain (see, e.g., \cite{Voit,Delft}).
The interwire interaction may transform the LL state existing in
isolated quantum wires into various phases of $2D$ quantum liquid. The most
drastic transformation is caused by the {\it interwire} tunneling $t_{\perp}$
in arrays of quantum wires with {\it intrawire} Coulomb repulsion.
This coupling constant rescales towards higher values for strong interaction
($g<1/2$), and the electrons in array transform into $2D$ Fermi liquid
\cite{Wen}.
The reason for this instability is the orthogonality catastrophe, i.e. the
infrared divergence in the low-energy excitation spectrum that accompanies
the interwire hopping processes.

Unlike interwire tunneling, the density-density or current-current
interwire interactions do not modify the low-energy behavior of quantum arrays
under certain conditions. In particular, it was shown recently
\cite{Luba00,Vica01,Luba01}
that an interaction of the type $W(n-n')$, which depends on the
distance between  wires $n$ and $n'$ but does not contain current
coordinates $x,x',$ imparts the properties of a {\it sliding phase} to 2D
array of 1D quantum wires. In this state an additional interwire coupling
leaves the  fixed-point action invariant under the "sliding" transformation
$\theta_n\to \theta_n+\alpha_n$ and $\pi_n \to \pi_n+\alpha^\prime_n$.
The contribution of interwire coupling reduces to a renormalization of the
parameters $v\to v(q_\perp)$, $g\to g(q_\perp)$ in the LL Hamiltonian
(\ref{D1}), where $q_\perp$ is a momentum perpendicular to the chain
orientation. Such LL structure can be interpreted as a quantum
analog of classical sliding phases of coupled $XY$
chains\cite{Hern}. Recently, it was found \cite{Sond} that a hierarchy of
quantum Hall states emerges in sliding phases when a quantizing magnetic field
is applied to an array. \\

In the present paper we concentrate on another aspect of the problem
of interacting quantum wires.  Instead of studying the conditions
under which the LL behavior is preserved in spite of interwire
interaction, we consider situations where the {\it dimensional
crossover} from $1D$ to $2D$ occurs.  In other words, we investigate
regimes, where the excitations in quantum array demonstrate either
$1D$ or $2D$ behavior in different parts of phase space.  The most
promising type of artificial structures where this effect may be
expected is a periodic $2D$ system of two arrays of parallel quantum
wires crossing each other at an angle $\varphi$.  We call it
"quantum crossbars" (QCB).  The square grids of this type were
considered in various physical context in early papers
\cite{Avr,Avi,Guinea,Castro,Kuzm}.  In Refs.\cite {Guinea,Castro} the
fragility of the LL state against interwire tunneling in the crossing
areas of QCB was studied.  It was found that a new periodicity
imposed by the interwire hopping term results in the appearance of a
low-energy cutoff $\Delta_l\sim \hbar v/a$ where $a$ is a period of
the quantum grid.  Below this energy, the system is "frozen" in its lowest
one-electron state.  As a result, the LL state remains robust against
orthogonality catastrophe, and the Fermi surface conserves its 1D
character in the corresponding parts of the $2D$ Brilllouin zone.  This
cutoff energy tends to zero at the points where the one-electron
energies for two perpendicular arrays $\epsilon_{k_{1}}$ and
$\epsilon_{k_{2}}$ become degenerate.  As a result, a dimensional
crossover from $1D$ to $2D$ Fermi surface (or from LL to FL behavior)
arises around the points $\epsilon_{F_{1}}=\epsilon_{F_{2}}$.\\

We study this dimensional crossover for Bose excitations (plasmons)
described by canonical variables $\theta,\pi$ in QCB. In order to
unravel the pertinent physics we consider a grid with {\it short-range
capacitive inter-wire interaction}.  This approximation seems natural
for $2D$ grids of carbon nanotubes \cite{Rueckes}, or artificially
fabricated bars of quantum wires with grid periods $a_{1,2}$ which
exceed the lattice spacing of a single wire or the diameter of a
nanotube.  It will be shown below that this interaction can be made
effectively weak.  Therefore, QCB retains the $1D$ LL character for
motion along the wires similarly to the case considered in Ref.
\cite{Luba01}.  At the same time, the boson mode propagation along
some resonant directions is also feasible.  This is essentially a $2D$
process in the $2D$ Brillouin zone of the reciprocal space. \\

\section{Quantum Crossbars: Basic notions}

\begin{figure}[htb]
\centering
\includegraphics[width=75mm,height=40mm,angle=0,]{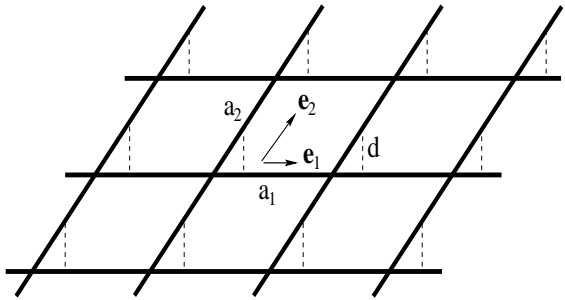}
\caption{$2D$ quantum bar formed by two
interacting arrays of parallel quantum wires. Here ${\bf e}_{1}, {\bf
e}_{2}$ are the unit vectors of the superlattice, ${a_{1}},{a_{2}}$
are the superlattice periods and $d$ is the vertical interarray
distance}
\label{Bar}
\end{figure}

A quantum crossbars may be defined as a $2D$ periodic grid, i.e. two
periodically crossed arrays of $1D$ quantum wires.  In fact these
arrays are placed on two parallel planes separated by an inter-plane
distance $d$ \cite{Rueckes}, but in this section we consider QCB as a
genuine $2D$ system.  We assume that all wires of the $j$-th array,
$j=1,2,$ have the same length $L_j,$ Fermi velocity $v_{j}$ and
Luttinger parameter $g_{j}.$ They are oriented along unit vectors
${\bf e}_{1,2}$ with an angle $\varphi$ between them.  Thus, the QCB
periods along these directions are $a_{1}$ and $a_{2},$ and the
corresponding QCB basic vectors are ${\bf a}_j=a_j{\bf e}_j$
(Fig.\ref{Bar}).  The interaction between the excitations in different
wires is assumed to be concentrated near the crossing points with
coordinates $n_1{\bf a}_1+n_2{\bf a}_2\equiv(n_1a_1,n_2a_2)$.  The
integers $n_j$ enumerate the wires within the $j$-th array.  Such
interaction imposes a superperiodicity on the energy spectrum of
initially one dimensional quantum wires, and the eigenstates of this
superlattice are characterized by a $2D$ quasimomentum ${\bf
q}=q_1{\bf g}_1+q_2{\bf g}_2 \equiv(q_1,q_2)$.  Here ${\bf g}_{1,2}$
are the unit vectors of the reciprocal superlattice satisfying the
standard orthogonality relations $({\bf e}_i\cdot {\bf
g}_j)=\delta_{ij}$.  The corresponding basic vectors of the reciprocal
superlattice have the form $(m_1Q_1, m_2Q_2)$, where $Q_j=2\pi/a_j$
and $m_{1,2}$ are integers.  \\

\begin{figure}[htb]
\centering
\includegraphics[width=75mm,height=80mm,angle=0,]{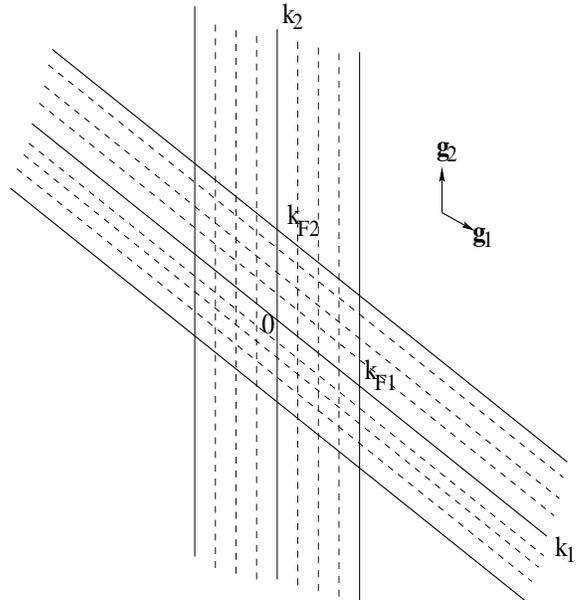}
\caption{Fermi surface of $2D$ metallic quantum
bar in the absence of charge transfer between wires.  ${\bf
g}_{1}, {\bf g}_{2}$ are the unit vectors of the reciprocal
superlattice} \label{FS}
\end{figure}

In conventional $2D$ systems, forbidden states in the inverse space
arise due to Bragg diffraction in a periodic potential, whereas the
whole plane is allowed for wave propagation in real space, at least
till the periodic potential is weak enough.  In strongly anisotropic
QCB, most of the real space is forbidden for electron and plasmon
propagation, whereas the Bragg conditions for the wave vectors are
still the same as in conventional $2D$ plane modulated by a periodic
potential.  The excitation motion in QCB is one-dimensional in major
part of the $2D$ plane, and the anisotropy in real space imposes
restrictions on the possible values of the $2D$ coordinates $x_{1},x_{2}$
(${\bf r}=x_{1}{\bf e}_{1}+x_{2}{\bf e}_{2}$).  At least one of them,
e.g., $x_2$ ($x_{1}$) should be an integer multiple of the
corresponding array period $a_2$ ($a_{1}$), so that the vector ${\bf
r}=(x_1,n_2 a_2)$ (${\bf r}=(n_1a_1,x_2)$) characterizes a point
with a $1D$ coordinate $x_1$ ($x_2$) lying at the $n_2$-th
($n_1$-th) wire of the first (second) array.\\

The $2D$ Brillouin zone of QCB is constructed as an extension of $1D$
Brillouin zones of two crossed arrays and subsequent folding of this
BZ in accordance with the $2D$ superstructure.  However, one cannot
resort to the standard basis of $2D$ plane waves when constructing an
eigenstate with a given wave vector ${\bf k}$ in the BZ because of the
kinematic restrictions mentioned above.  Even in {\it non-interacting}
arrays of quantum wires the $2D$ basis is two sets of $1D$ waves.
These are $1D$ excitations propagating along each wire of array
$1$ characterized by a unit vector $k_1{\bf g}_1$ with a phase shift
$a_2k_2$ between adjacent wires, and the same for array 2.  The
states of equal energy obtained by means of this procedure form
straight lines in the $2D$ BZ. Respectively, the Fermi sea is not a
circle with radius $k_F$ like in the case of free $2D$ gas, but a
cross in the $k$ plane bounded by these four lines \cite{Guinea} (see
Fig.\ref{FS}).  \\

\begin{figure}[htb]
\centering
\includegraphics[width=75mm,height=75mm,angle=0,]{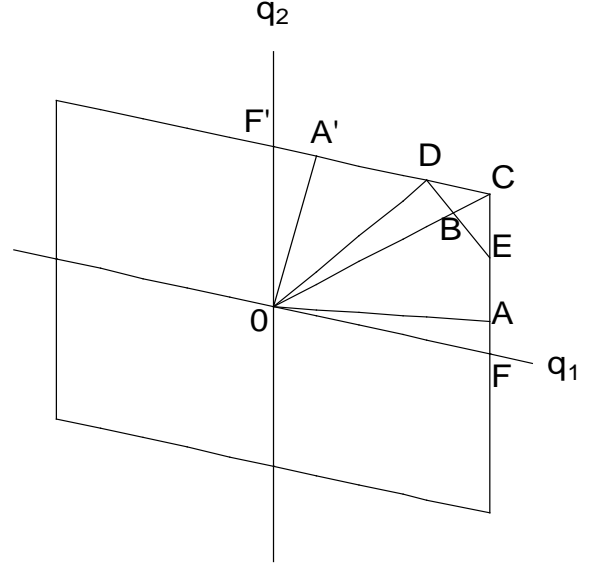}
\caption{Two dimensional Brillouin zone of QB.
Four polygonal lines along which the dispersion of Bose
excitations is calculated in Section V are marked by $AOA'$, $FCF'$,
and $ODE$.} \label{BZ}
\end{figure}

Due to the weak inter-wire interaction, the excitations in the $2D$ BZ
depicted in Fig.\ref{BZ} acquire two-dimensionality
characterized by the quasimomentum ${\bf q}=(q_1,q_2)$.  However, in
case of interaction, the $2D$ waves constructed from the $1D$
plane waves in accordance with the above procedure form an appropriate
basis for the description of elementary excitations in QCB, in close
analogy with the nearly free electron approximation in conventional
crystalline lattices.  It is easy to believe that the inter-wire
interaction does not completely destroy the above quasimomentum
classification of eigenstates, and the $2D$ reconstruction of the
spectrum may be described in terms of wave mixing similarly to the
standard Bragg diffraction in a weak periodic potential.  Moreover,
the classification of eigenstates of non-interacting crossed arrays of
$1D$ wires ("empty superlattice") may be effectively used for the
classification of energy bands in a real QB superlattice.  Our next
task is to construct a complete $2D$ basis for this empty
superlattice.\\

Excitations in a given wire are
described as plane waves $L^{-1/2}\exp (ikx)$ with wave number $k$ and
initial dispersion law $\omega(k)=v|k|$ (the array number is
temporarily omitted ).  Each excitation in an ``empty superchain'' is
described by its quasi wavenumber $q$ and a band number $s$
($s=1,2,\ldots$).  Its wave function has the Bloch-type structure,
\begin{equation}
    \psi_{s,q}(x)=\frac{1}{\sqrt{L}}
    e^{iqx}
    u_{s,q}(x).
    \label{WaveFunc}
\end{equation}
We confine ourselves with the first BZ of a superchain, $|q|\leq Q/2,$
where the Bloch amplitude $u_{s,q}$  and dispersion law $\omega_{s}$
have the following form:
\begin{eqnarray}
    u_{s,q}(x) & = &
    \exp
    \left\{
    iQx(-1)^{s-1}\left[\frac{s}{2}\right]
    {\mbox{ sign }}q
    \right \},
    \label{WaveFunc1}\\
    \omega_{s}(q) & = &  vQ\left(
    \left[\frac{s}{2}\right]
    +(-1)^{s-1}\frac{|q|}{Q}
    \right).
    \label{Disp1}
\end{eqnarray}
To write down these formulas for a specific array, one should add the
array index $j$ to the wave function $\psi$, Bloch amplitude $u$,
coordinate $x$, quasimomentum $q$, periods $a$ and $Q$ of the $1D$
lattice in real and reciprocal space.\\

The $2D$ basis of periodic Bloch functions for an empty superlattice
is constructed in terms of $1D$ Bloch functions
(\ref{WaveFunc}), (\ref{WaveFunc1})
\begin{equation}
{\Psi}_{s,s',{\bf q}}({\bf r})={\psi}_{1,s,q_{1}}(x_{1})
            {\psi}_{2,s',q_{2}}(x_{2}).
            \label{Psi}
\end{equation}
Here the $2D$ quasimomentum ${\bf q}=(q_{1}, q_{2})$ belongs to the
first BZ, $|q_{j}|\leq Q_j/2.$ The corresponding eigenfrequencies are
\begin{equation}
    \omega_{ss'}({\bf q})=
    \omega_{1,s}(q_{1})+\omega_{2,s'}(q_{2}).
    \label{Omega}
\end{equation}
We will use this basis in the next section when constructing the
excitation spectrum of QB within the reduced band scheme.\\

The full Hamiltonian of the QB is,
\begin{equation}
  H =  {H}_{1} + {H}_{2} +  H_{{int}},
  \label{TotHam1}
\end{equation}
where ${H}_{j}$ describes the $1D$ boson field characterised by the
parameters $v_j, g_j$ in the $j$-th array (see eq.  (\ref{D1})), and
$H_{int}$ is the interwire interaction.  One may neglect inter-wire
tunneling and restrict oneself by the capacitive interaction only,
provided the vertical distance $d$ between the wires is substantially
larger than the screening radii $r_{j}$ within the wires.  Then
\begin{eqnarray}
H_{{int}} & = & V_{0}\sum\limits_{{n}_{1},{n}_{2}}
            \int dx_1 dx_2
            \Phi\left
        (\frac{x_1-n_1a_1}{r_{1}},\frac{n_2a_2-x_2}{r_{2}}\right)\times
    \nonumber\\
        &&\partial_{x_1}\theta_1(x_1,n_2a_2)
         \partial_{x_2}\theta_2(n_1a_1,x_2).
\label{H_int_1}
\end{eqnarray}

It stems from the Coulomb interaction
between intrawire charge fluctuations within the crossing area.
The size of intrawire fluctuations is
determined by $r_{j}$.
The couplig strength is
$
V_0=2e^{2}/d,
$
and the function $\Phi(\xi_{1},\xi_{2})$ is
\begin{equation}
    \Phi(\xi_{1},\xi_{2})=
        \frac
    {
    \displaystyle{
    \zeta_1(\xi_{1})
    \zeta_2(\xi_{2})
    }}
    {\displaystyle{
    \sqrt{1+|{\bf r}_{12}|^2/d^2
    }}}\approx \zeta_1(\xi_{1})
    \zeta_2(\xi_{2})
    ,
        \label{F}
\end{equation}
provided $|{\bf r}_{12}|^2/d^2\ll 1$.
Thus the  interaction is separable in this limit.

The above approximation looks realistic for QCB fabricated from carbon
nanotubes \cite{Rueckes}.  In this case the Coulomb interaction is
screened at a distance of the order of the nanotube
radius\cite{Sasaki} $R_{0},$ therefore $r_{1,2}\sim R_{0}.$ The
minimal radius of a single-walled carbon nanotube is about
$R_{0}=0.35\div 0.4 nm$ (see \cite{Louie}).  The vertical distance $d$
in artificially produced nanotube networks is estimated as $d\approx
2$nm \cite{Rueckes}.  Therefore the ratio
$r_{0}^{2}/d^{2}\approx{0.04}$ is really small.

In the quasimomentum representation
(\ref{Psi},\ref{WaveFunc},\ref{WaveFunc1})
the full Hamiltonian (\ref{TotHam1}) acquires the form,
\begin{eqnarray}
H & = & \frac{{\hbar}{v}{g}}{2}\displaystyle{
                      \sum_{j=1}^{2}
                      \sum_{s,{\bf q}}
                      }
                      {\pi}_{js{\bf q}}^{+}
                      {\pi}_{js{\bf q}}  +\nonumber\\
          &&\frac{\hbar}{2vg}\displaystyle{
          \sum_{jj'=1}^{2}
                 \sum_{s,s',{\bf q}}
                 }
                 {W}_{jsj's'{\bf q}}
                 {\theta}_{j s {\bf q}}^{+}
                 {\theta}_{j's'{\bf q}},
                 \label{TotHam2}
\end{eqnarray}
where $v=\sqrt{v_1v_2}$, $g=\sqrt{g_1g_2}$, while
$\sqrt{vg/v_{j}g_{j}}{\theta}_{js{\bf q}}$ and
$\sqrt{v_{j}g_{j}/vg}{\pi}_{js{\bf q}}$
 are the Fourier components of the boson fields ${\theta}_{j}$ and
${\pi}_{j}$.
The matrix elements for interwire coupling are given by:
\begin{eqnarray*}
{W}_{jsj's'{\bf{q}}} & = &
    {\omega}_{j s }(q_{j} )
    {\omega}_{j's'}(q_{j'})
    \left[
         {\delta}_{jj'}{\delta}_{ss'}+
%    \right.\nonumber\\ & & \left.
         {\phi}_{jsj's'{\bf{q}}}
         \left(
              1-{\delta}_{jj'}
         \right)
    \right].
\end{eqnarray*}
Here $\omega_{js}(q_{j})$ are eigenfrequencies (\ref{Disp1}) of the
``unperturbed'' $1D$ mode pertaining to an array $j$.  The
coefficients
\begin{eqnarray}
{\phi}_{1s2s'{\bf{q}}} = \phi(-1)^{s+s'}
                       \mbox{sign}{(q_{1}q_{2})}
                       {\Phi}_{1s2s'{\bf{q}}},\nonumber\\
                       \phi =
                       \frac{gV_{0}r_{0}^{2}}{{\hbar}va},\ \ \
               r_0=\sqrt{r_{1}r_{2}}, \ \ \
               a=\sqrt{a_1a_2},
                       \label{phi}
\end{eqnarray}
are proportional to the dimensionless Fourier component of the
interaction strengths
\begin{eqnarray}
{\Phi}_{1s2s'{\bf{q}}} & = &
  \int d{\xi}_{1}d{\xi}_{2}{\Phi}({\xi}_{1},{\xi}_{2})
  e^{-i(r_{1}q_1\xi_1+r_{2}q_2\xi_2)}\times\nonumber\\
    && u_{1,s, q_1}^{*}(r_{1}\xi_1)
  u_{2,s',q_2}^{*}(r_{2}\xi_2).
  \label{Phi}
\end{eqnarray}

The Hamiltonian (\ref{TotHam2}) describes a system of coupled harmonic
oscillators, and can be {\em exactly} diagonalized.  The
diagonalization procedure is cumbersome in the general case due to
mixing of states belonging to different bands and arrays.  However, in
the case of separable interwire potential (\ref{F}) one easily comes
to a compact secular equation for the eigenfrequencies of QCB:
\begin{equation}
 F_{1q_1}(\omega^2)F_{2q_2}(\omega^2)=\frac{1}{\varepsilon},
 \label{secul-eq}
\end{equation}
where
\begin{eqnarray}
 F_{jq}(\omega^2) & = & \frac{r_{j}}{a_j}\sum\limits_{s}
 \frac{\phi_{js}^{2}(q)\omega_{js}^{2}(q)}
      {\omega_{js}^{2}(q)-\omega^2},
 \label{F_j}\\
  \phi_{js}(q) & = & (-1)^s\mbox{sign}(q)\int d\xi
  \zeta_{j}(\xi) e^{ir_{j}q\xi}u_{jsq}(r_{j}\xi), \nonumber
\end{eqnarray}
and the dimensionless coupling constant $\varepsilon$ can be written as
\begin{equation}
    \varepsilon=\left(\phi\frac{a}{r_0}\right)^{2}=
    \left(\frac{gV_0r_0}{{\hbar}v}\right)^{2}=
\left(\frac{2R_{0}}{d}\frac{ge^{2}}{\hbar v}\right)^{2}.
    \label{epsilon}
\end{equation}
For nanotube QCB, the first factor within parentheses is about
$0.35.$ The second one, that is nothing but the corresponding QCB
``fine structure'' constant, can be estimated as $0.9$ (we used the
values of $g=1/3$ and $v=8\times 10^{7}$cm/sec, see Ref. \cite {Egger}).
Therefore $\varepsilon\approx 0.1,$ and the coupling is really weak.

\section{Energy spectrum}

Due to weakness of the interaction, the systematics of unperturbed
plasmon levels and states is grossly conserved, at least in the low
energy region corresponding to the first few bands.  This means that
perturbed eigenstates could be described by the same quantum numbers
as the unperturbed ones.  The interband mixing is significant only
along the high symmetry directions in the first BZ (BZ boundaries and
lines $q_{j}=0$).  In zeroth approximation with respect to the weak
interaction, these lines are determined by the Bragg conditions.
Inter-array mixing within the same energy band is strong only for
waves with quasimomenta close to the resonant lines in the BZ. In
zeroth approximation with respect to the interaction, these lines are
determined by the conditions
$\omega^{2}_{1s}(q_{1})=\omega^{2}_{2s'}(q_{2})$ with all possible
positive integers $s,s'$.  In the rest of the BZ, the initial
systematics can be used. \\

\begin{figure}[htb]
\centering
\includegraphics[width=80mm,height=95mm,angle=0,]{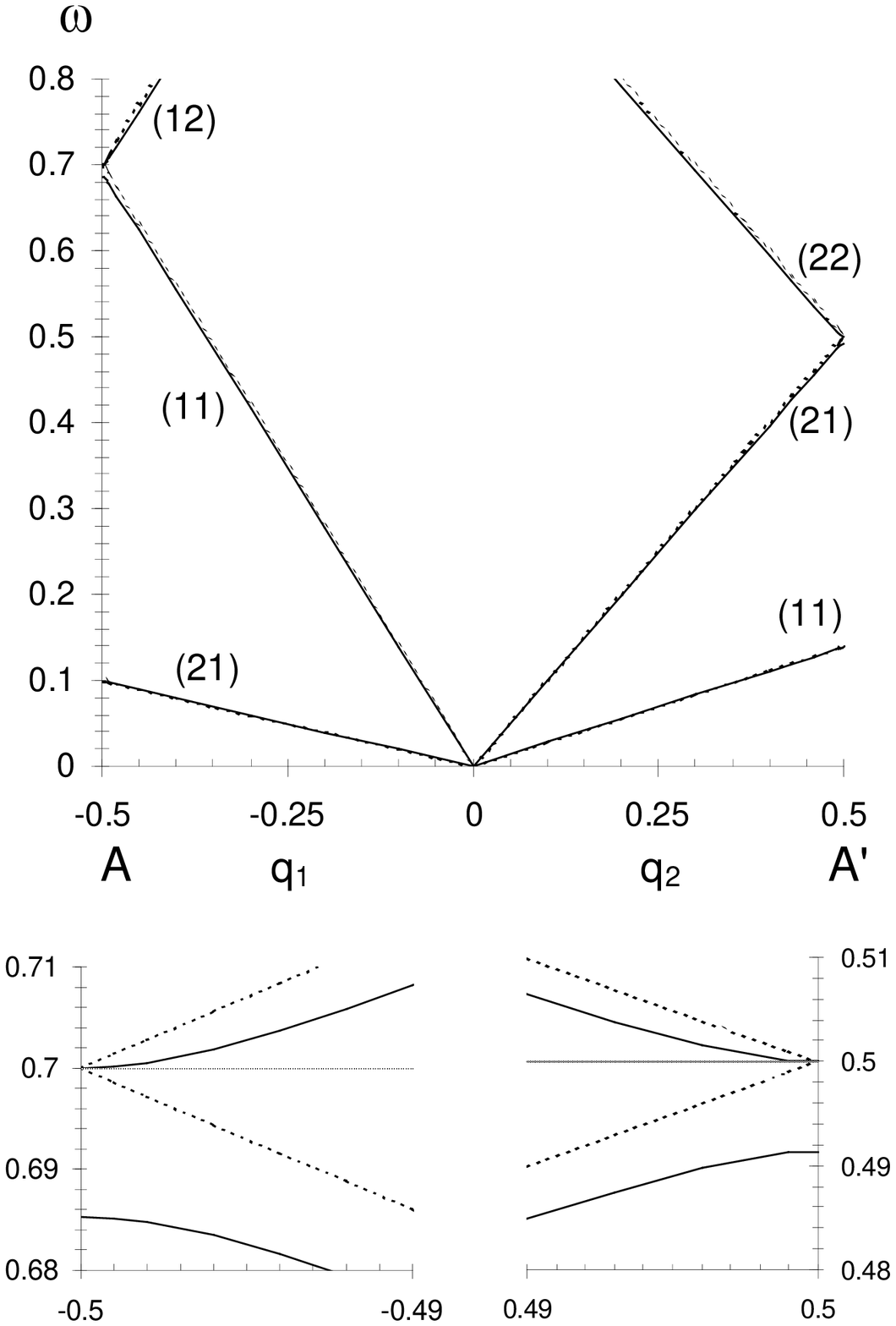}
\caption{The energy spectrum of QCB (solid lines) and
noninteracting arrays (dashed lines) for quasimomenta at the line
$AOA'$ of Fig.\ref{BZ} ($q_2=0.2q_1$ along line $AO$ and
$q_1=0.2q_2$ along line $OA'$). Insets: Zoomed vicinity of the
point $q_{1}/Q_1=0.5$, $\omega=0.7$ (left side); and the point
$q_{2}/Q_2=0.5$, $\omega=0.5$ (right side).} \label{sp-off}
\end{figure}

The three next figures illustrate the main features of the excitation
spectrum.  In Fig.\ref{sp-off} the dispersion curves, corresponding to
quasi momenta changing along the line $AOA'$ of Fig.\ref{BZ} are
plotted in comparison with those for non interacting arrays.  (In all
figures within this section we use units $\hbar=Q_2=v_2=1$, and
$v_1Q_1=1.4$).  In what follows we use $(j,s)$ notations for the
unperturbed boson propagating along the $j$-th array in the $s$-th
band.  Then the lowest curve in the left part of Fig.\ref{sp-off}
(line $AO$ in Fig.\ref{BZ}) is, in fact, the slightly renormalized
dispersion of a $(2,1)$ plasmon, the middle curve describes $(1,1)$
plasmon, and the upper curve is the dispersion of a $(1,2)$ plasmon.
The fourth frequency, corresponding to a $(2,2)$ plasmon, is far above
and is not displayed in this part of the figure.  The right part of
Fig.\ref{sp-off} describes $(1,1)$ plasmon (lowest curve), $(2,1)$
plasmon (middle curve) and $(2,2)$ plasmon (upper curve).  It is seen
that the dispersion remains linear along the whole line $AOA'$ except
at a nearest vicinity of the BZ boundary (see insets in
Fig.\ref{sp-off}).  It is clearly seen that the plasmon preserve their
1D character along these lines, and small deviation from linearity is
observed only near the boundaries of the BZ. The interband
hybridization gaps for bosons propagating along the $j$-th array can
be estimated as $ {\Delta\omega}^j_{12}\sim {v_jQ_j}\varepsilon
r_{0}/a$.  \\

\begin{figure}[htb]
\centering
\includegraphics[width=70mm,height=60mm,angle=0]{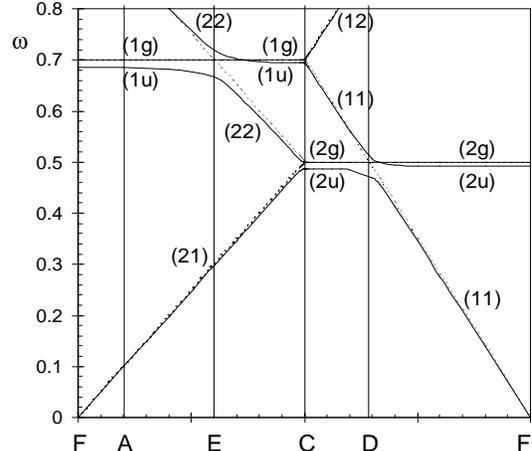}
\caption{The energy spectrum of QCB (solid lines)
and noninteracting arrays (dashed lines) for quasimomenta at the
boundary of the BZ (line $FCF'$ in the Fig.\ref{BZ})}
\label{sp-bound}
\end{figure}

More pronounced effects of wave mixing are seen in Fig.
\ref{sp-bound} where the dispersion curves corresponding to the line
FCF' (Fig.\ref{BZ}) along the boundary of the BZ are plotted.  Again,
the dispersion laws retain their $1D$ character along the major part
of the boundary.  The interaction opens the gap in the $1D$ bands for
arrays 1 and 2 along the lines $FC$ and $CF'$ respectively.  Odd (u)
and even (g) combinations of two waves are formed as a result of wave
mixing.  Strong 2D effects are observed around the points $D$ and $E$.
As a result we observe the {\it dimensional crossover} $1D \to 2D$
when moving along the boundary of the Z.\\

The strongest wave-mixing effects are observed along the line $ODE$ in
Fig.\ref{BZ}.  Here the plasmons belonging to arrays 1 and 2 are mixed
along the whole line.  They form odd and even combinations but the
dispersion is nearly linear everywhere except at the vicinity of the
points $D$ and $E$ where {\it three-wave mixing} takes place.  In
a square QCB the strong wave-mixing occurs in the vicinity of the
diagonals of the BZ \cite{Kuzm}.\\

\begin{figure}[htb]
\centering
\includegraphics[width=70mm,height=60mm,angle=0,]{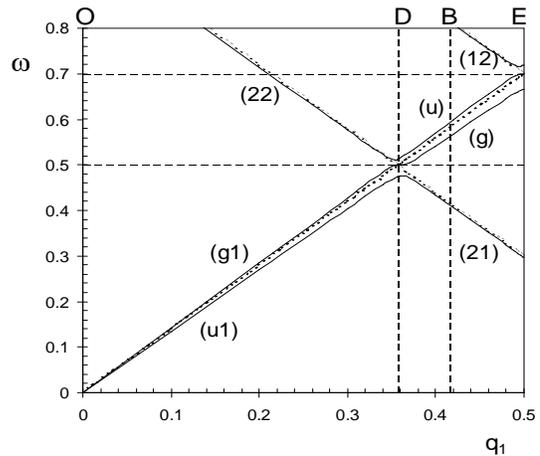}
\caption{The energy spectrum of QCB (solid lines) and
noninteracting arrays (dashed lines) for quasimomenta at the resonant
line of the BZ (line $ODE$ in Fig.\ref{BZ})}
\label{sp-diag}
\end{figure}

The low-energy part of the spectrum along most part of the line $OD$
is described by the secular equation
\begin{equation}
  \prod\limits_{j=1}^{2}
  \left(
       \frac{\varphi_j^2(q_j)\omega^2}
            {\omega_j^2(q_j)-\omega^2}+
       F_j
  \right)
  =\frac{1}{\varepsilon},
  \label{sq-eq}
\end{equation}
which follows from the general equation(\ref{F_j}).  Its solution
gives two nearly linear plasmon bands which conserve their LL
character in spite of the $2D$ wave mixing.  Just this solution is
described by eq.  (3.10) of Ref.  \cite{Luba01}.  So, our exact
procedure confirms the conclusion of renormalization approach of this
paper that the sliding LL phase may exist in two-dimensional QCB, and
the inter-array density-density interaction is irrelevant for the LL
fixed point.\\

\begin{figure}[htb]
\centering
\includegraphics[width=85mm,height=85mm,angle=0,]{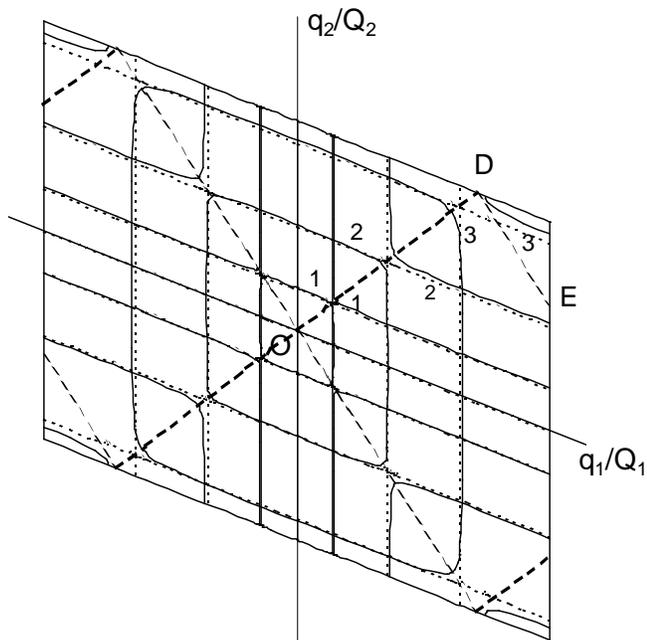}
\caption{The lines of equal frequency for QCB (solid
lines) and noninteracting arrays (dashed lines).  The lines
$1,2,3$ correspond to the frequencies
$\omega_{1}=0.1$, $\omega_{2}=0.25$, $\omega_{3}=0.45$.}
\label{IsoEn1}
\end{figure}

\begin{figure}[htb]
\centering
\includegraphics[width=85mm,height=90mm,angle=0,]{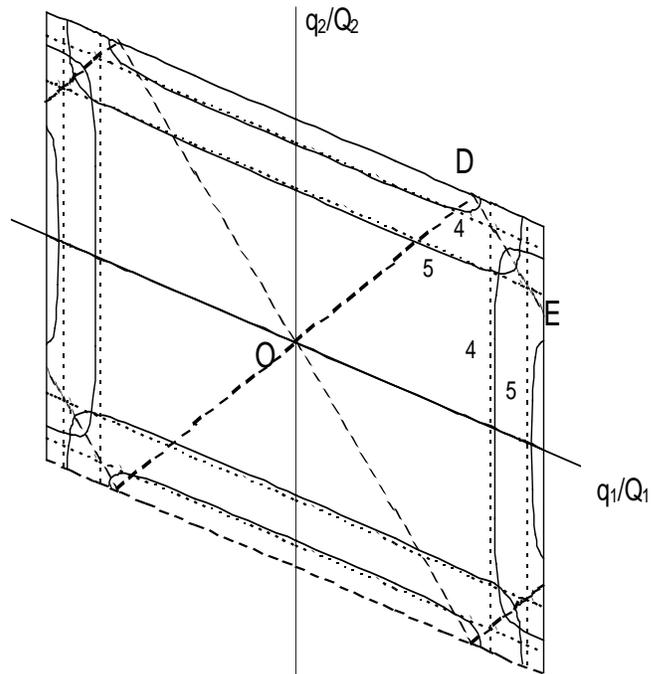}
\caption{The lines of equal frequency for QCB (solid
lines) and noninteracting arrays (dashed lines). The lines
$4,5$ in the lower panel correspond to the frequencies
$\omega_{4}=0.55$, $\omega_{5}=0.65$}
\label{IsoEn2}
\end{figure}

Finally we show the lines of equal frequency for Bose excitations
(Figs. \ref{IsoEn1}.\ref{IsoEn2}). These lines should be compared with
the Fermi surface "cross" shown in Fig. \ref{FS}. Their rounding
near the broken line $ODE$ is a manifestation of $1D \to 2D$ crossover.
Similar rounding of the 1D Fermi surface due to inter-array tunneling
was discussed in Ref. \cite{Castro}.

\section{Correlations and observables}
The correlation functions of QCB in the infrared limit are usually
discussed in
a framework of the theory of sliding LL phases \cite{Luba01}. These are
the Drude peak in the optical conductivity,
$\sigma(\omega)=D(T)\delta(\omega)$, the power-low temperature dependence
of resistivity, and the crossover from isotropic to anisotropic
conductivity at a certain length scale,
when the current is inserted at a point on
array 1 and extracted at another point on array 2.
All these features are reproduced by our exact solution which
generates the LL thermodynamics and transport as an intrinsic property of
QCB Hamiltonian (\ref{TotHam2}).

In this section we discuss in brief the correlation properties which
allow one to reveal specific $2D$ features of QCB at {\it finite
frequency and momentum.} One of the main effects specific for a QCB is
the appearance of non-zero transverse momentum--momentum correlation
function.  In space-time coordinates $({\bf{x}},t)$ it reads,
\begin{equation}
G_{12}({\bf x},t) =
           \left\langle \left[
                             {\pi}_{1}(x_1,0;t),
                             {\pi}_{2}(0,x_2;0)
           \right] \right\rangle.
\end{equation}
This function describes the momentum response at the point
$(0,x_{2})$ of the second array at time $t$ caused by
initial ($t=0$) perturbation at the point $(x_{1},0)$ of the first
array. Standard calculations lead to
the following expression,
\begin{eqnarray}
& {G}_{12}({\bf{x}};t) =
                     \displaystyle{
                     -i\frac{V_0r_0^2}{4{\pi}^2{\hbar}}
                     \int\limits_{-\infty}^{\infty} dk_1 dk_2
                     }
                     {\phi}_1({k_1}){\phi}_2({k_2})k_1k_2
                     \times &
                     \nonumber \\
                     & \times
                     \sin(k_1x_1)
                     \sin(k_2x_2)
                     \displaystyle{
                     \frac{v_2k_2\sin(v_2k_2t)-
                           v_1k_1\sin(v_1k_1t)}
                          {v_2^2k_2^2-v_1^2k_1^2},
                     } &
              \label{spat_trans}
\end{eqnarray}
where ${\phi}_{j}(k)$ is the Fourier component (\ref{F_j})
written in the extended BZ.\\

\begin{figure}[htb]
\centering
\includegraphics[width=75mm,height=75mm,angle=0,]{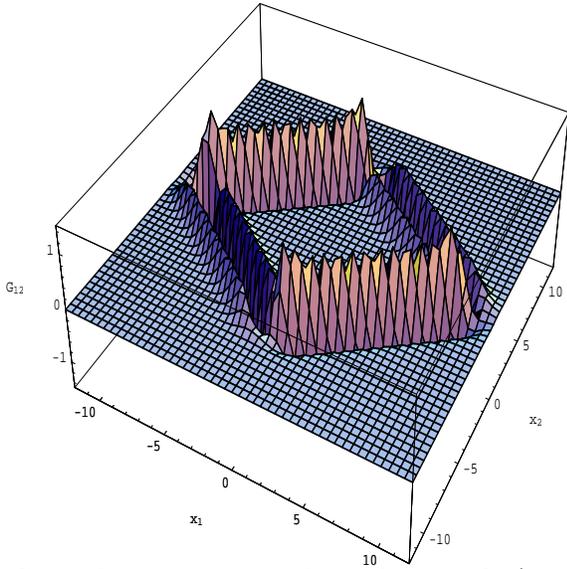}
\caption{The transverse correlation function $G_{12}(x_1,x_2;t)$
for $r_0=1$ and $vt=10$}
\label{GF12}
\end{figure}

This correlator is shown in Fig. \ref{GF12}.  Here the non-zero
response corresponds to the peak located at the line determined by the
obvious kinematic condition $|x_{1}|+|x_{2}|=vt.$ The finiteness of
the interaction radius slightly spreads this peak and changes its
profile.\\

\begin{figure}[htb]
\centering
\includegraphics[width=70mm,height=100mm,angle=0,]{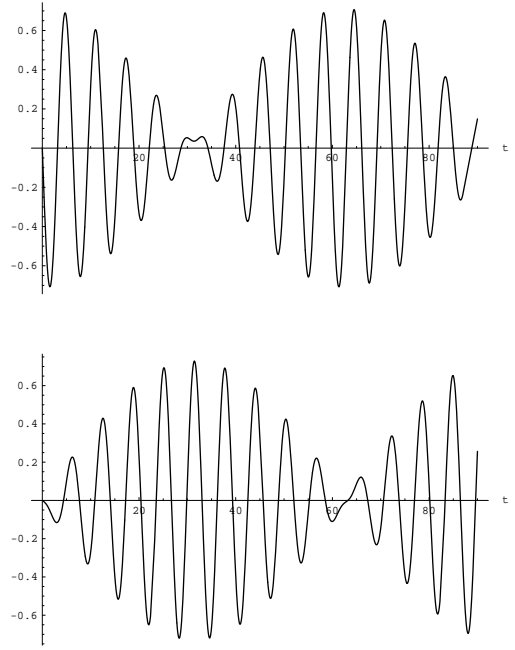}
\caption{Periodic energy exchange between arrays (``Rabi oscillations'')}
\label{RO2}
\end{figure}

Further manifestation of the $2D$ character of QCB
is related to a periodic energy transfer between the two
arrays of wires. Consider an initial perturbation which, in the
system of {\it non}-interacting arrays, excites a plane
wave propagating within the first array along the ${\bf e_{1}}$
direction,
\begin{eqnarray}
  \langle
  {\theta}_{1}(x_1,n_2a_2;t)
  \rangle & = &
  \frac{{\rho}_{0}}{\sqrt{2}|q_1|}
  \sin(q_1x_1+q_2n_2a_2-v_1|q_1|t),\nonumber\\
\langle
  {\theta}_{2}(n_{1}a_1,x_{2};t)
  \rangle & = & 0,
\end{eqnarray}
($\rho_{0}$ is the charge density amplitude).  If the wave vector
${\bf q},$ satisfying the condition $|{\bf{q}}|<<Q_{1,2}/2,$ is
not close to the resonant line of the first BZ, weak interwire
interaction $\phi=\varepsilon{r}_{0}/{a}$ slightly changes the
$\langle{\theta}_{1}\rangle$ component and leads to the appearance
of a small $\langle{\theta}_{2}\rangle\sim\phi$ component.  But
for ${\bf q}$ lying on the resonant line
($v_1|q_1|=v_2|q_2|\equiv\omega_{\bf{q}}$), both components within
the main approximation have the same order of magnitude
\begin{eqnarray}
   {\theta}_{1}(x_1,n_2a;t) & = &
   \frac{\rho_0}{\sqrt{2}|q_1|}
   \cos\left(
                 \frac{1}{2}
                 {\phi}_{\bf{q}}
                 {\omega}_{{\bf{q}}}t
            \right)\times\nonumber\\
   &&\sin(q_1x_1+q_2n_2a_2-{\omega}_{{\bf{q}}}t),
   \label{theta1d}
\end{eqnarray}
\begin{eqnarray}
   {\theta}_{2}(n_1a_1,x_2;t) & = &
   \frac{\rho_0}{\sqrt{2}|q_1|}
   \sin\left(
            \frac{1}{2}
            {\phi}_{\bf{q}}
            {\omega}_{{\bf{q}}}t
   \right)\times\nonumber\\
      &&\cos(q_1n_1a_1+q_2x_2-{\omega}_{{\bf{q}}}t).
   \label{theta2d}
\end{eqnarray}
Here $\phi_{\bf{q}}\equiv \phi_{1121{\bf{q}}}$ (see Eq.(\ref{phi}).
This corresponds to a $2D$ propagation of a plane wave with wave vector
${\bf q },$ {\it modulated} by a ``slow'' frequency $\sim\phi\omega.$
As a result, an energy is periodically transferred from one array to
another during a long period $T\sim (\phi\omega)^{-1}$ (see
Fig.\ref{RO2}).  These peculiar ``Rabi oscillations'' may be considered
as one of the fingerprints of the physics exposed in QCB systems.

Interarray interaction affects also the optical conductivity $\sigma(\omega)$.
It was shown in Ref.\cite{Kuzm} that as a result a transverse component
$\sigma_{\perp}(\omega)$ appears.

%%%%%%%%%%%RO!!!%%%%%%%%%%%%%%%%%
\section{Conclusion}
We have shown that the bosonization procedure may be applied to the
Hamiltonian of 2D quantum grids at least in the first few Brillouin
zones.  The energy spectrum of QCB shows the characteristic properties
of LL at $|q|,\omega\to 0,$ but at finite ${\bf q},$ the density and
momentum waves may have either $1D$ or $2D$ character depending on the
direction of the wave vector.  Due to interwire interaction,
unperturbed states, propagating along the two arrays are always mixed,
and transverse components of correlation functions do not vanish.  For
quasi-momenta near the diagonal of the BZ, such mixing is strong, and
the transverse correlators possess specific dynamical properties.

%%%%%%%%%%%%%%%%%%%%%

\end{document}